\documentclass[12pt]{iopart}

\usepackage{hyperref}
\usepackage{graphicx}
\usepackage{epstopdf} 
\usepackage{dcolumn}
\usepackage{bm}
\usepackage{wasysym}
\usepackage{amssymb}
\usepackage{color}

\begin{document}
\bibliographystyle{unsrt}

\title{Parallel and perpendicular turbulence correlation length in the TJ-II Stellarator}

\author{B.Ph.~van Milligen$^1$, A. Lopez Fraguas$^1$, M.A.~Pedrosa$^1$, C. Hidalgo$^1$, A.~Mart\'in de Aguilera$^1$, E. Ascas\'ibar$^1$}
 \address{$^1$Asociaci{\'o}n EURATOM-CIEMAT para Fusi{\'o}n, Avda.~Complutense 40, 28040 Madrid, Spain}
 \date{\today}

\begin{abstract}
Long range correlations were measured using two remote reciprocating Langmuir probe systems at TJ-II.
The influence of the rotational transform on the correlation was studied by scanning the magnetic configuration.
A simple drift wave correlation model, assuming an exponential decay of the correlation with different correlation lengths in the directions parallel and perpendicular to the field lines, was found to describe the observations well at low densities.
The experiment was repeated at gradually higher densities, and an additional correlation was detected at a critical value of the density. In accordance with previous work, this additional correlation was ascribed to zonal flows associated with a confinement transition.
Thus, the total long range correlation is found to be a sum of the drift wave and zonal flow contributions.  
\end{abstract}

\maketitle

\section{Introduction}

The understanding of the three-dimensional structure of turbulence in magnetically confined plasmas is an important issue in research towards fusion as an alternative energy source. In the past, much effort has been directed towards this end, using a wide range of experimental techniques~\cite{Zweben:2007}. The tried technique of measuring turbulent fluctuations using Langmuir probes permits determining the correlation structure between spatially separated probe pins in sufficiently cool regions of the plasma~\cite{Ritz:1988,Bengtson:1998, Thomsen:2001,Xu:2009}.
Recently, this technique was used to determine the three-dimensional structure of turbulence in a small-size, weakly heated device, using a dense array of probes~\cite{Birkenmeyer:2012}.

In the present work, we pursue a similar goal in a medium-size, strongly heated confinement device, the TJ-II stellarator.
TJ-II possesses two reciprocating Langmuir probe systems, located at toroidal angles $\phi = 38.2 ^\circ$ (probe D) and $\phi = 194.5 ^\circ$ (probe B). In the past, the signals measured by these probes were found to show significant correlation when the two probes were located approximately on the same flux surface, depending on plasma conditions~\cite{Pedrosa:2008,Pedrosa:2010}. The correlation was found to be strongest for floating potential measurements, and less but still significant for ion saturation current measurements.
The correlation was observed to vary systematically with the local value of the rotational transform $\iota/2\pi$, which was interpreted as being related to the presence (or not) of rational surfaces at the probe location.

Correlation between the signals measured by spatially separated probes may arise due to various causes. 
Any global mode (MHD or other) affecting the measured signals will produce significant correlation due to the global nature of the mode~\cite{Yu:2003}.
Global modes are easily recognized due to their coherent oscillatory nature, producing a clear peak in the spectrum, while MHD modes are recognized by their magnetic footprint (coherent oscillations detected by Mirnov magnetic field pick-up coils).

In the absence of global modes, the probe signals may still be correlated due to, e.g., the preferential confinement of supra-thermal electrons at rational surfaces, or the local reduction of Neoclassical viscosity~\cite{Velasco:2012}, or the presence of zonal flows associated with flux surfaces (generated by the turbulence via the Reynolds Stress mechanism)~\cite{Diamond:2005}.

However, correlation may also arise due to a more basic mechanism.
Drift wave turbulence has an elongated structure along the magnetic field lines~\cite{Chen:1984,Birkenmeyer:2012}, related to the virtually unimpeded movement of electrons along field lines, while their perpendicular motion is strongly restrained.
Assuming such turbulent structures are continuously being produced and have a finite lifetime, 
their elongated structure would imply a certain correlation between two Langmuir probes, depending on the (parallel and perpendicular) size (or wavenumber $k$) of the structure and the location of the two probes with respect to the field line geometry.

In this work, we report measurements of probe correlation in low density plasmas without significant MHD activity or zonal flows, and compare the observed correlation with a simple prediction based on the vacuum magnetic field configuration. 
This allows making a rough determination of the parallel and perpendicular correlation lengths (or $k$'s) inside the plasma.
Furthermore, we study the impact of a spontaneous confinement transition on the correlation.

\clearpage
\section{Methods} 

\begin{figure}\centering
  \includegraphics[trim=0 0 0 0,clip=,width=14cm]{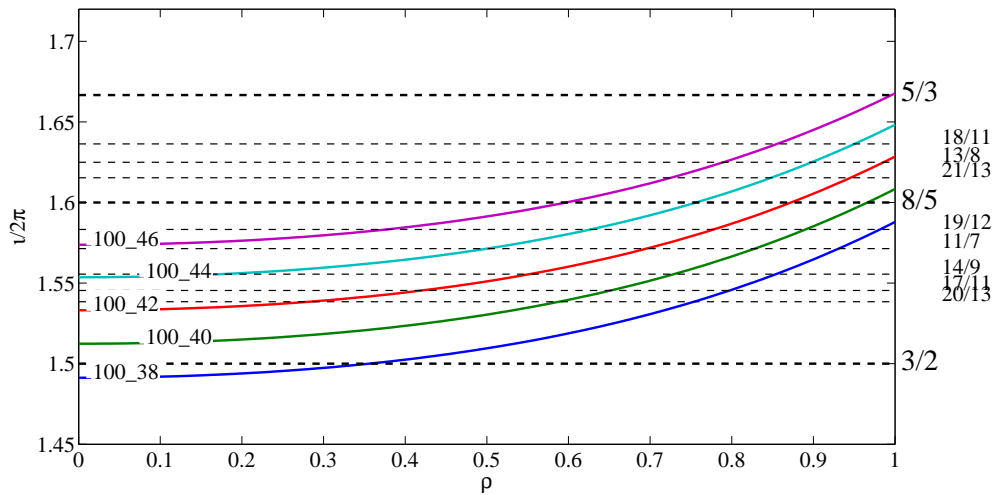}
\caption{\label{iota}The rotational transform, $\iota/2\pi$, for various relevant configurations. Some rational values of $\iota/2\pi$ are indicated.}
\end{figure}

We performed a series of dynamic configuration scans by varying the rotational transform at the edge, $\iota(1)/2\pi$, slowly and linearly in time (by modulating the external coil currents), while using plasma current control to keep currents inside the plasma small ($|I_p| < 0.5$ kA). 
Thus, the magnetic configuration is very tightly controlled, particularly in the external part of the plasma, $\rho =r/a > 0.8$, where the influence of plasma currents on the magnetic configuration is small.
As mentioned, we inserted two probes into the plasma at approximately the same magnetic surface but at remote toroidal positions.
By gradually varying the rotational transform, the connection length between the two probes was modified, which in turn should affect the measured correlation between probe signals.
Table~\ref{table1} shows the 4 scans that were performed, while Fig.~\ref{iota} shows the corresponding $\iota/2\pi$ profiles.
The shape of the $\iota/2\pi$ profiles is very similar except for an offset.

\begin{table}[htdp]
\caption{The four configuration scans discussed here. The scans occur between $t=1100$ ms and $t=1250$ ms.}
\begin{center}
\begin{tabular}{|c|c|c|c|c|}
\hline
Scan number & Start configuration & End configuration & Start $\iota(1)/2\pi$ & End $\iota(1)/2\pi$ \\
\hline
1 & 100\_46\_65 & 101\_42\_64 & 1.672 & 1.630 \\
2 & 101\_42\_64 & 100\_46\_65 & 1.630 & 1.672 \\
3 & 101\_42\_64 & 101\_38\_62 & 1.630 & 1.593 \\
4 & 101\_38\_62 & 101\_42\_64 & 1.593 & 1.630 \\
\hline
\end{tabular}
\end{center}
\label{table1}
\end{table}%

A simple method for relating the observed correlation to the magnetic field structure was established in the pioneering work of \cite{Ritz:1988}.
However, in the present case, the range of variation of the rotational transform is not sufficiently broad to apply this method directly. 
In particular, it was not possible to achieve a situation in which the two probes are directly linked by a magnetic field line (within one toroidal turn).
Hence, we present a novel procedure to extract the geometric correlation parameters (correlation lengths) using a least squares method.

Given the known vacuum magnetic field geometry for each value of $\iota(1)/2\pi$, the geometric distances $d_\perp$ and $L_{||}$ are computed, as follows.
Starting from the nominal position of the reference probe (probe D), the magnetic field line is followed both in the positive and negative toroidal direction. 
Each toroidal turn of the field line is assigned a label, $i$.
In a given toroidal turn of the magnetic field line, the point of closest approximation to probe B is found. 
The distance of closest approximation is $d_\perp(i)$. 
The corresponding distance to the reference probe D, measured along the field line, is called $L_{||}(i)$.
As noted, the distances $d_\perp(i)$ and $L_{||}(i)$ depend on the configuration (i.e., on $\iota(1)/2\pi$).
Thus, for each value of $\iota(1)/2\pi$ in the scans discussed below, $d_\perp(i)$ and $L_{||}(i)$ are recomputed.

For each value of $\iota(1)/2\pi$, the expected correlation is estimated by  assuming that the correlation decays exponentially in the directions perpendicular and parallel to the field line, and considering the contributions from all close approximations of the field line:
\begin{equation}\label{C}
C_{DB} = A \sum_{i}{\exp \left ( -\frac{d_\perp(i)^2}{\lambda_\perp^2} -\frac{L_{||}(i)^2}{\lambda_{||}^2} \right )}
\end{equation}
where $A$ is a proportionality constant, and $i$ enumerates the close approximations.
Thus, in this simple model, the correlation between probes D and B is a function of only three parameters ($A$, $\lambda_\perp$, and $\lambda_{||}$), assumed fixed during each $\iota/2\pi$ scan for simplicity.

\clearpage
\section{Results}

The plasma was heated by Electron Cyclotron Resonant Heating (nominal power: $2 \times 250$ kW). The density was kept low and approximately constant in the time interval of the $\iota/2\pi$ scans, $1100 \le t \le 1250$ ms, as shown in Table \ref{table2}. Thus, both the shot to shot and the temporal variation of $\overline{n_e}$ was kept below 5 \%.

\begin{table}[htdp]
\caption{Line averaged density values (mean value and standard deviation) for the 4 scans}
\begin{center}
\begin{tabular}{|c|c|c|c|}
\hline
Scan & Nr. shots & $\overline{n_e} (10^{19}$ m$^{-3}$) & $\Delta \overline{n_e} (10^{19}$ m$^{-3})$ \\ 
\hline
1 & 10 & 0.42 & 0.015 \\
2 & 6 & 0.40 & 0.015 \\
3 & 10 & 0.40 & 0.02 \\
4 & 10 & 0.43 & 0.015 \\
\hline
\end{tabular}
\end{center}
\label{table2}
\end{table}%

\begin{figure}\centering
  \includegraphics[trim=0 0 0 0,clip=,width=10cm]{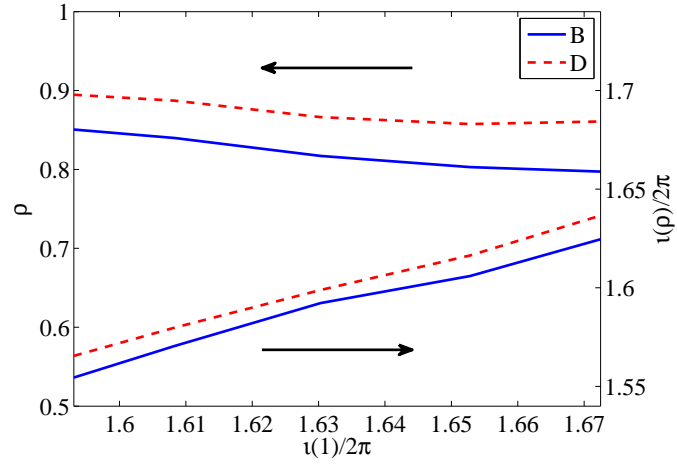}
\caption{\label{iotascan}The normalized radius, $\rho$, and rotational transform, $\iota(\rho)/2\pi$, at the nominal positions of the two probes (B and D) versus the rotational transform at $\rho=1$, $\iota(1)/2\pi$. The $x$-axis range corresponds to the full configuration scan range considered here (cf.~Table \ref{table1} and Fig.~\ref{iota}).}
\end{figure}

In these experiments, the two reciprocating Langmuir probes D ($\phi = 38.2 ^\circ$) and B ($\phi = 194.5 ^\circ$) were inserted a certain distance into the plasma so as to be located approximately on the same flux surface. The nominal position of the probes corresponds to their true position with an error of less than 1 cm.
Fig.~\ref{iotascan} shows the nominal normalized radius and the value of the rotational transform at the nominal position of the two Langmuir probes versus the edge rotational transform, for the vacuum configurations. The radial separation of the probes was below 6\% of the minor radius, i.e., of the same order as the positioning error of the probes (1 cm).
In the plasmas considered here, the ion Larmor radius is of the order of 1 mm.

\subsection{Coherence and spectra}

\begin{figure}\centering
  \includegraphics[trim=0 0 0 0,clip=,width=7.5cm]{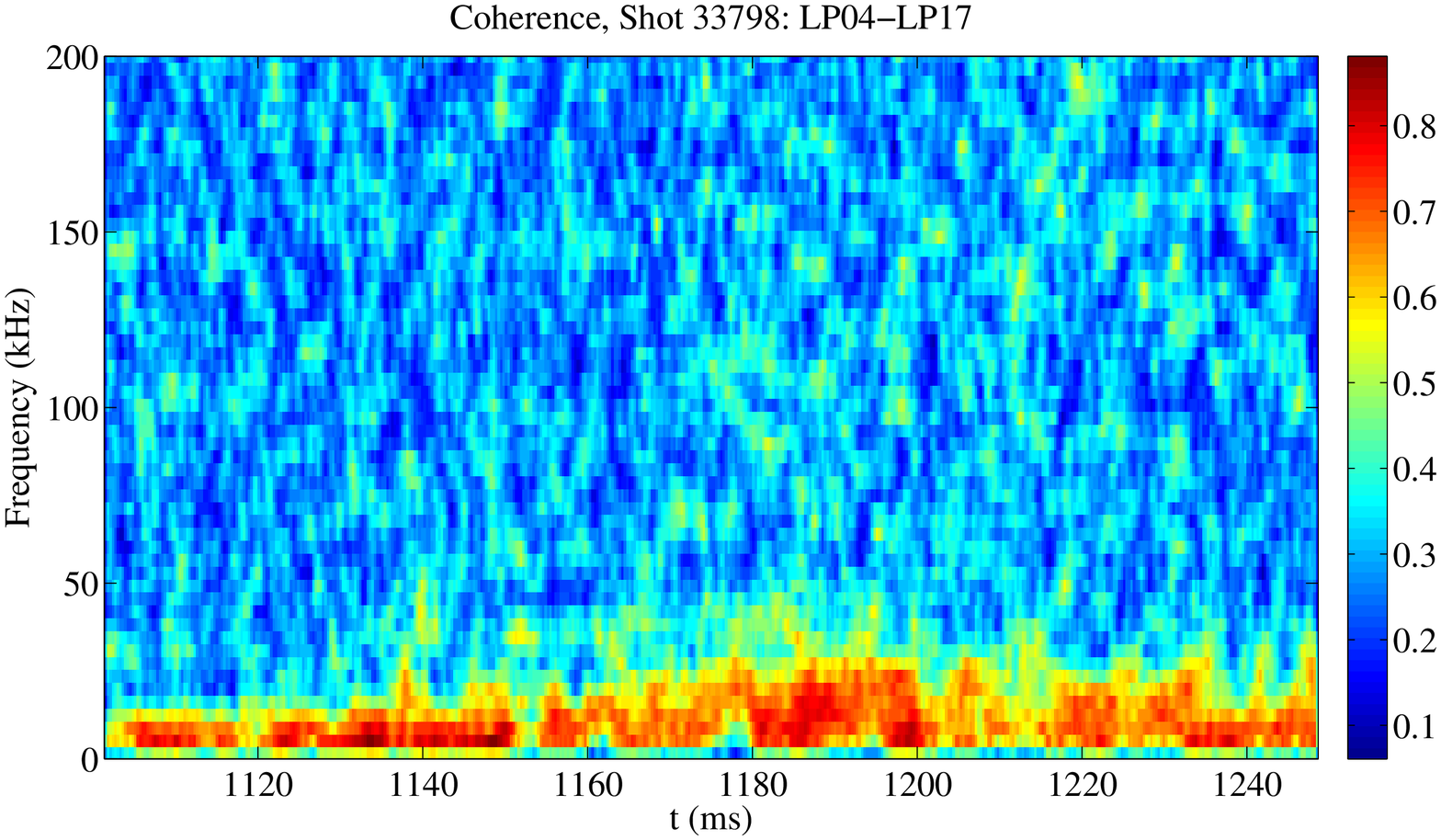}
  \includegraphics[trim=0 0 0 0,clip=,width=7.5cm]{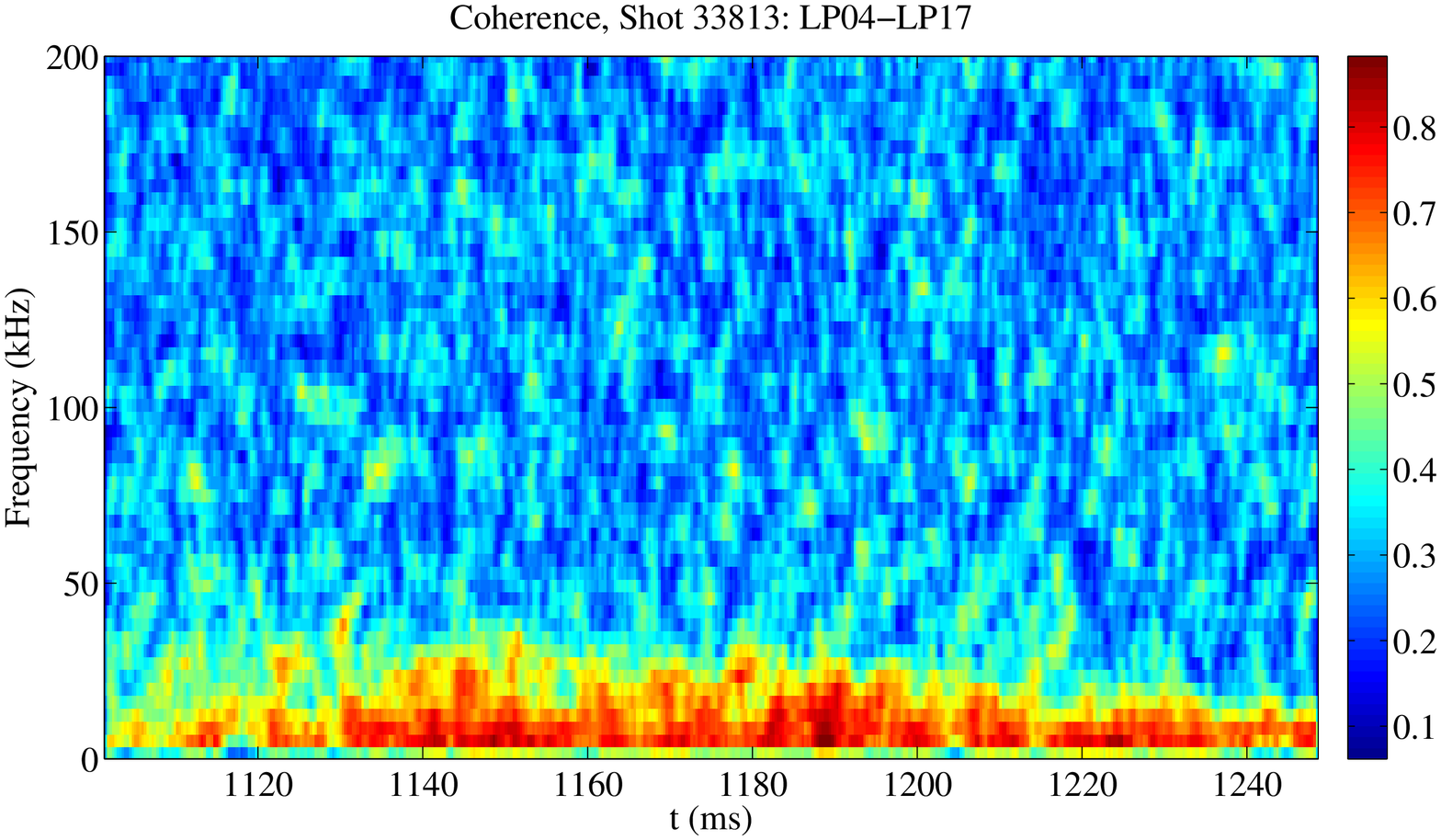}\\
  \includegraphics[trim=0 0 0 0,clip=,width=7.5cm]{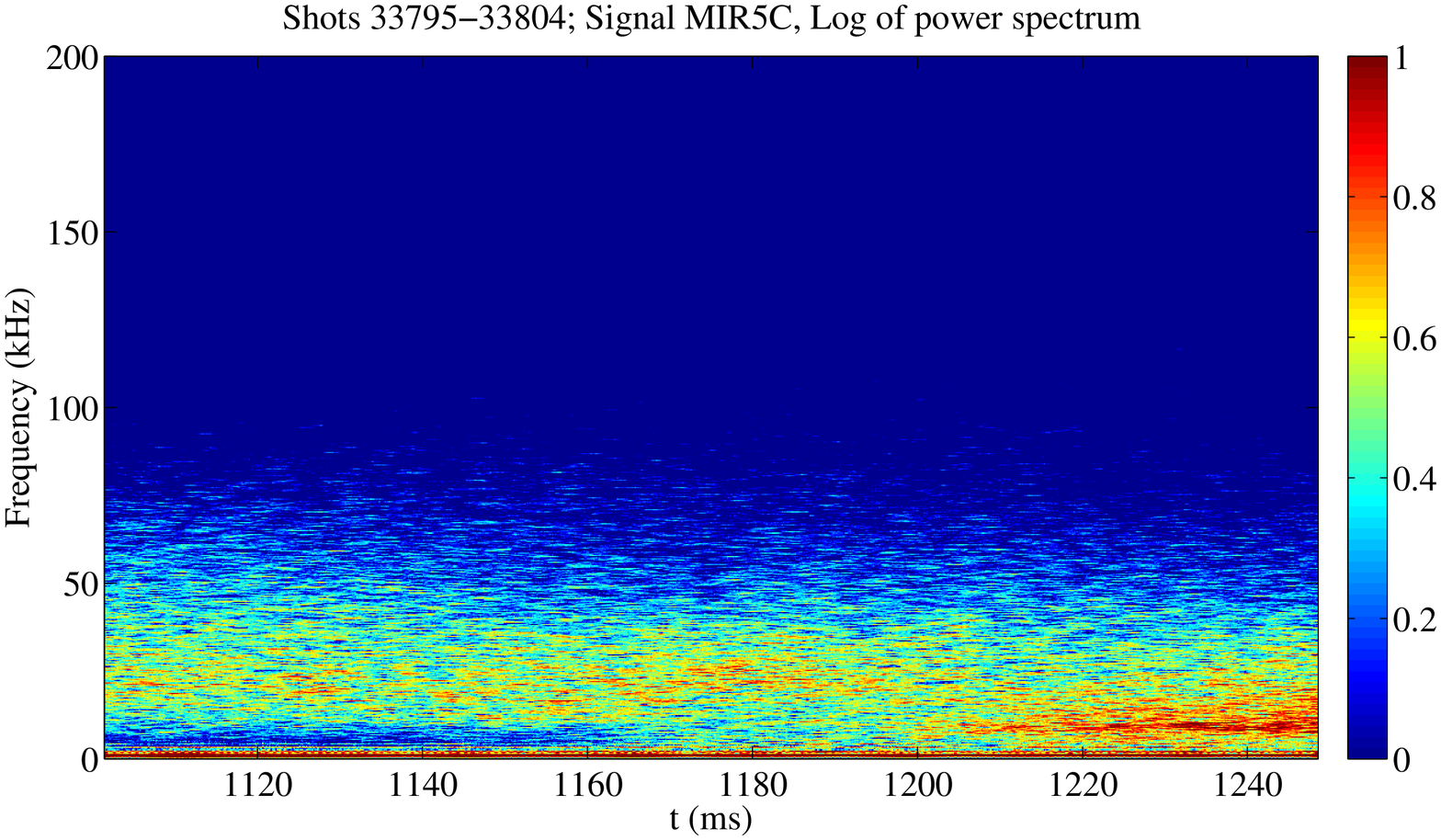}
  \includegraphics[trim=0 0 0 0,clip=,width=7.5cm]{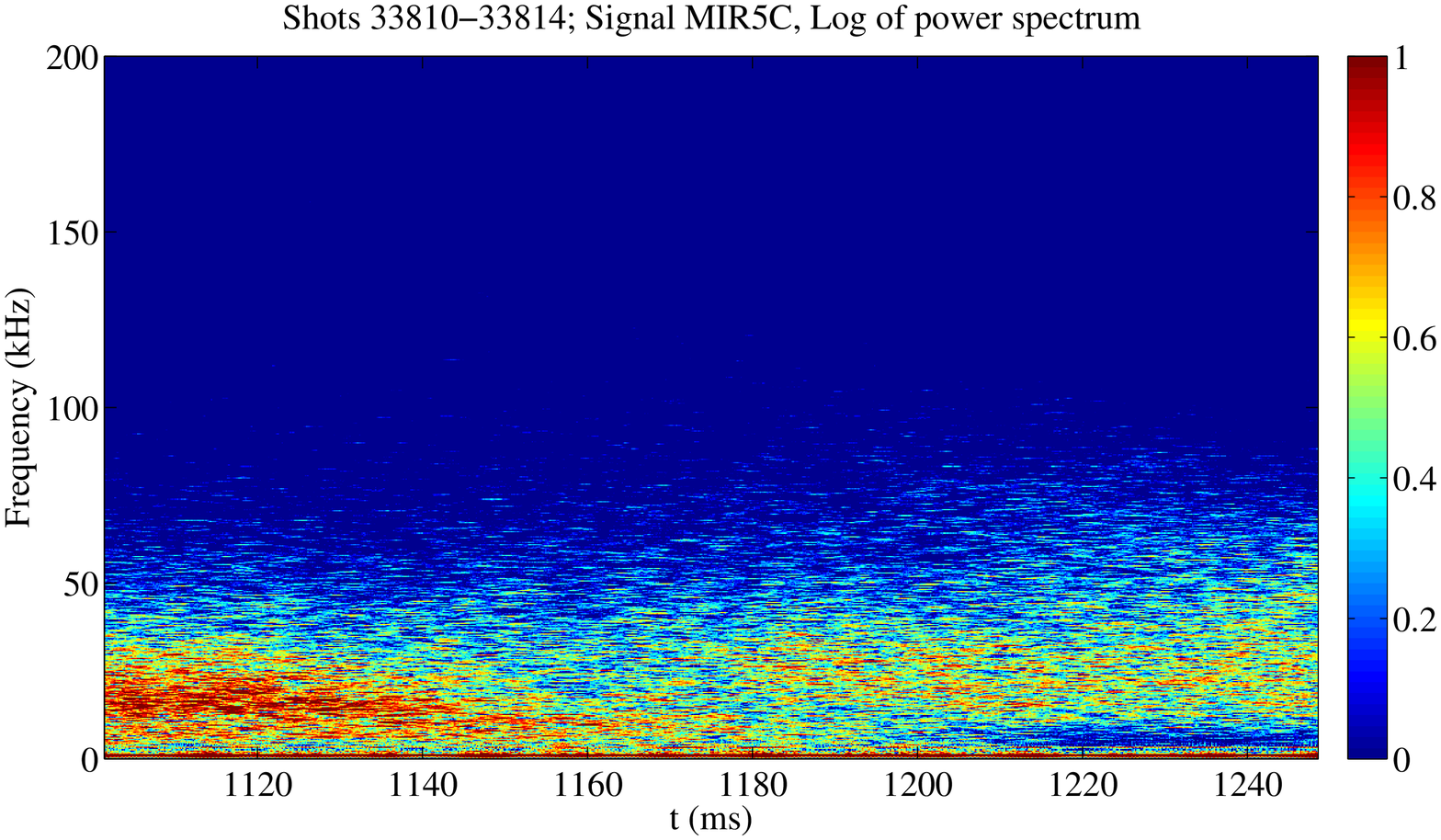}
\caption{\label{spec_scan1-2}Spectra for scans 1 and 2.
Top: cross coherence of floating potential signals measured by probes D and B (left: a discharge from scan 1; right: a discharge from scan 2).
Bottom: mean spectra of a Mirnov coil (mean taken over 10 and 6 repetitive discharges, for scans 1 and 2, respectively).
Color scale: logarithm, base 10, of the spectral power.}
\end{figure}

Fig.~\ref{spec_scan1-2} shows the cross coherence of floating potential signals measured by probes D and B, and spectra of a magnetic field Mirnov pick-up coil in the same time interval, for scans 1 and 2.
While there is some magnetic activity, much of it occurs at a higher frequency than the frequencies corresponding to the probe coherence. Furthermore, the interval of high coherence spans nearly the whole time window, whereas the magnetic activity is concentrated in a specific time window.
The magnetic spectrum is not very coherent, indicating there are no significant global MHD modes, although there is some minor magnetic activity at $t > 1200$ ms (scan 1) or $t<1150$ ms (scan 2), corresponding to the lower end of the $\iota/2\pi$ scan, when the 8/5 rational is located in the edge region around $\rho \simeq 0.85$ (cf.~Fig.~\ref{iota}).

\begin{figure}\centering
  \includegraphics[trim=0 0 0 0,clip=,width=7.5cm]{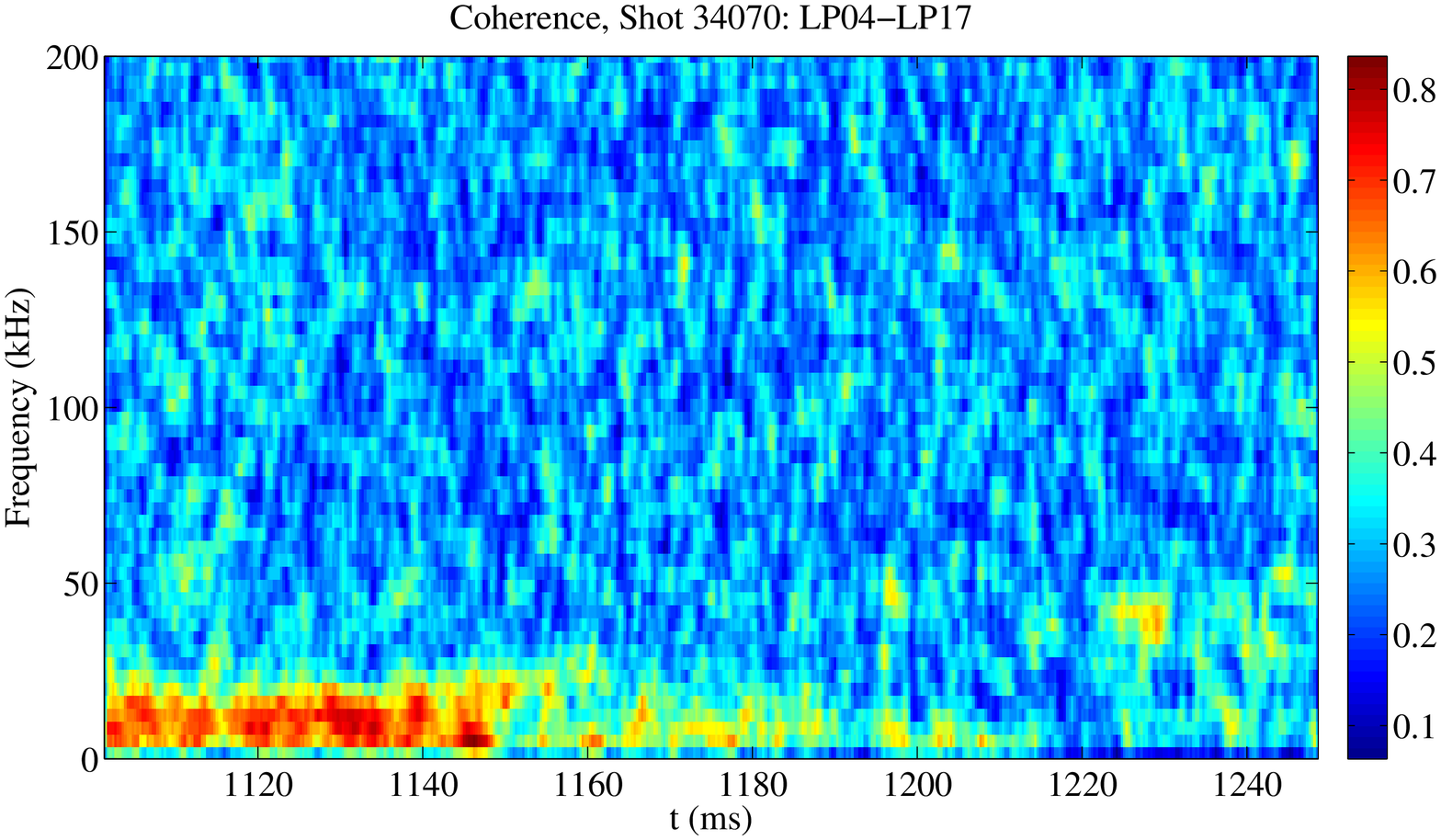}
  \includegraphics[trim=0 0 0 0,clip=,width=7.5cm]{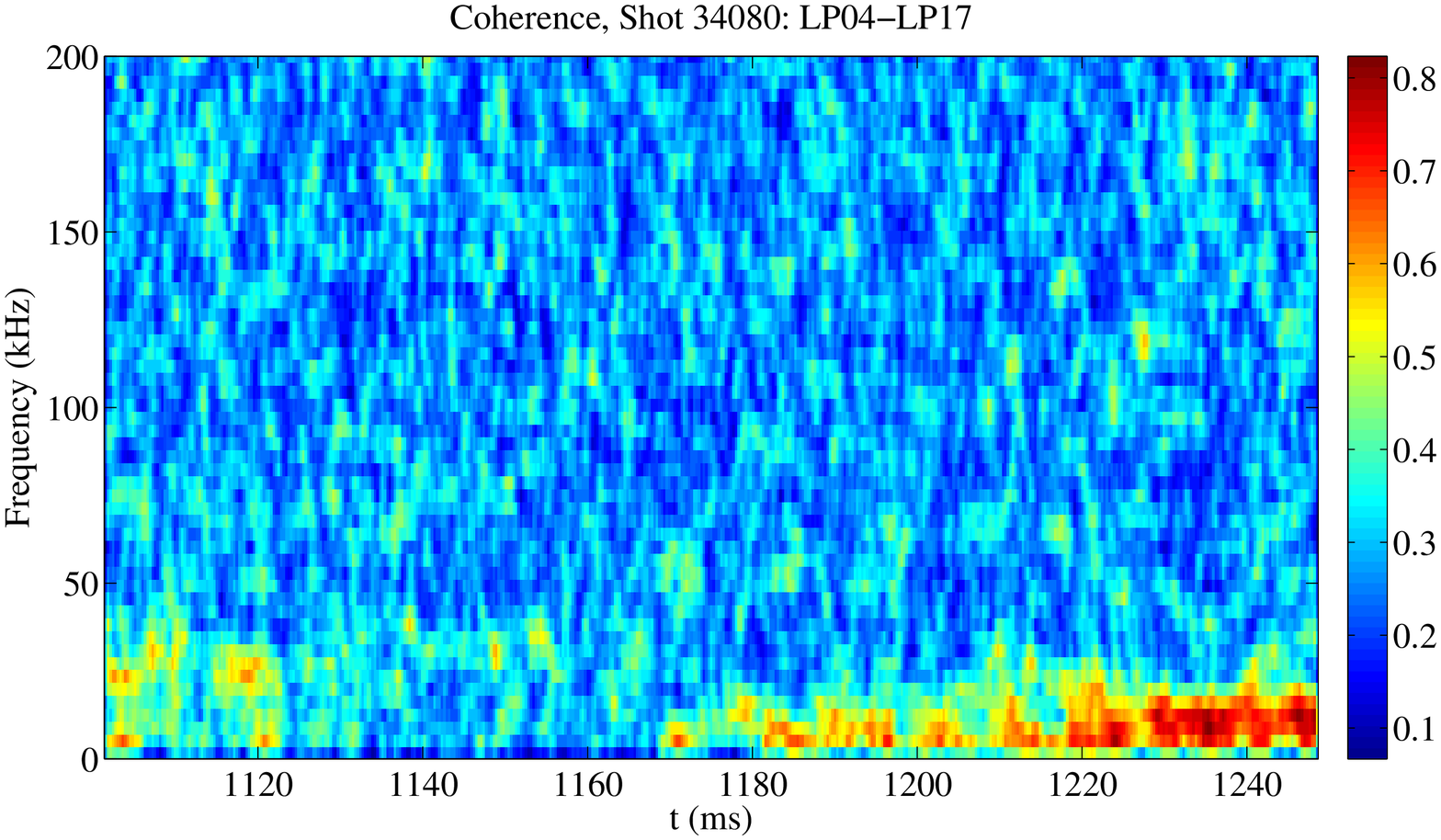}\\
  \includegraphics[trim=0 0 0 0,clip=,width=7.5cm]{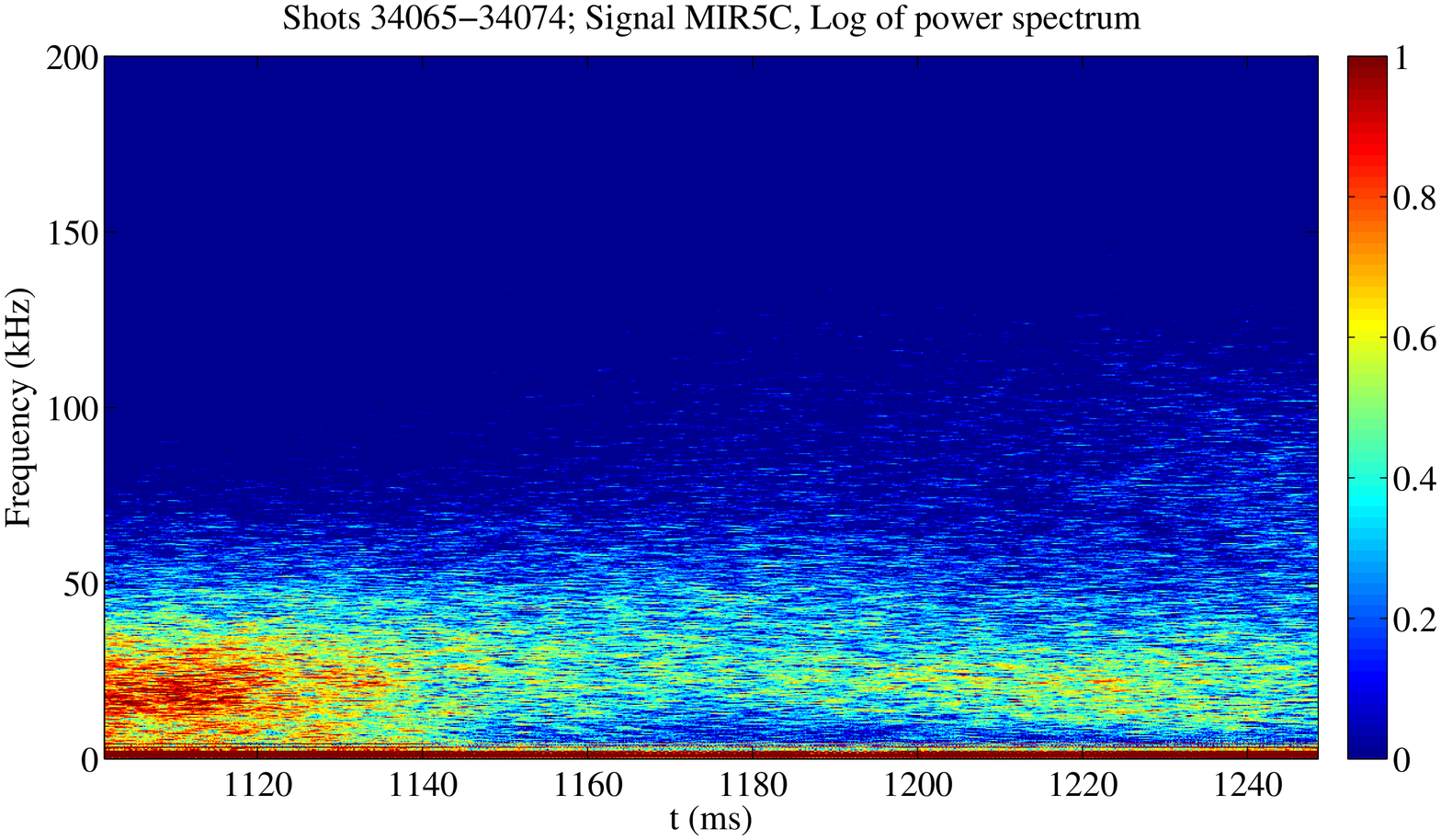}
  \includegraphics[trim=0 0 0 0,clip=,width=7.5cm]{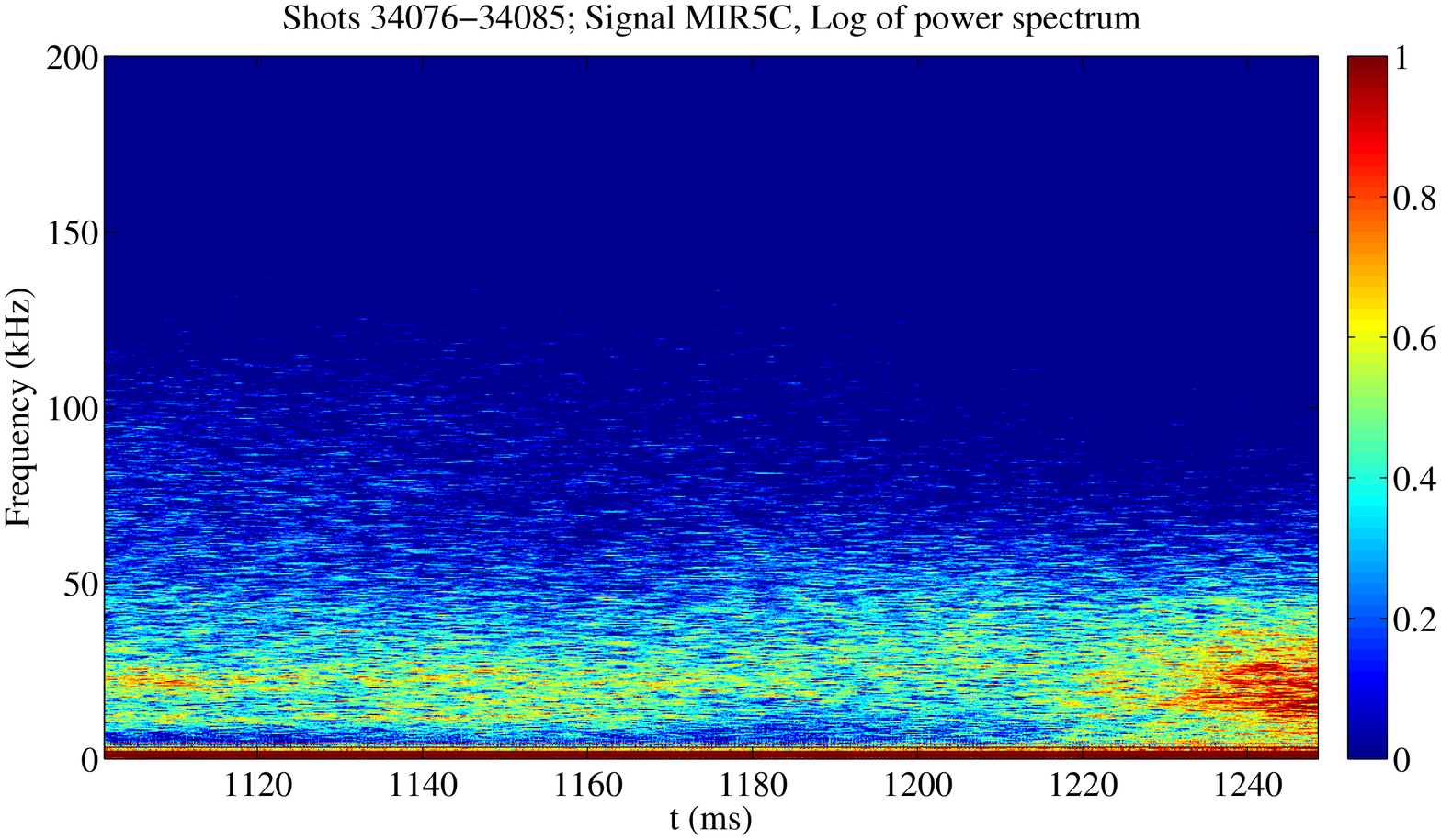}
\caption{\label{spec_scan3-4}Spectra for scans 3 and 4.
Top: cross coherence of floating potential signals measured by probes D and B (left: a discharge from scan 3; right: a discharge from scan 4).
Bottom: mean spectra of a Mirnov coil (mean taken over 10 repetitive discharges for scans 3 and 4, respectively).
Color scale: logarithm, base 10, of the spectral power.}
\end{figure}

Fig.~\ref{spec_scan3-4} shows a similar picture for scans 3 and 4.
Both the coherence and the magnetic spectra are relatively high for $t<1150$ ms (scan 3, downward) or $t>1220$ ms (scan 4, upward), when the 8/5 rational is located in the edge region around $\rho \simeq 0.9$ (cf.~Fig.~\ref{iota}). 
Again, magnetic activity is not very coherent.

Summarizing the spectral information from all four scans, no significant coherent mode activity was detected.
There is some minor magnetic activity when $\iota/2\pi$ is close to its central value in the full scan range (the red profile labelled 100\_42 in Fig.~\ref{iota}), when the 8/5 rational surface is located between $0.8 < \rho < 0.9$.
Thus, it is not expected that global MHD modes contribute significantly to the observed correlation (see below), although a minor contribution cannot be ruled out.

\clearpage
\subsection{Long range correlation at low density}

\begin{figure}\centering
  \includegraphics[trim=0 0 0 0,clip=,width=7.5cm]{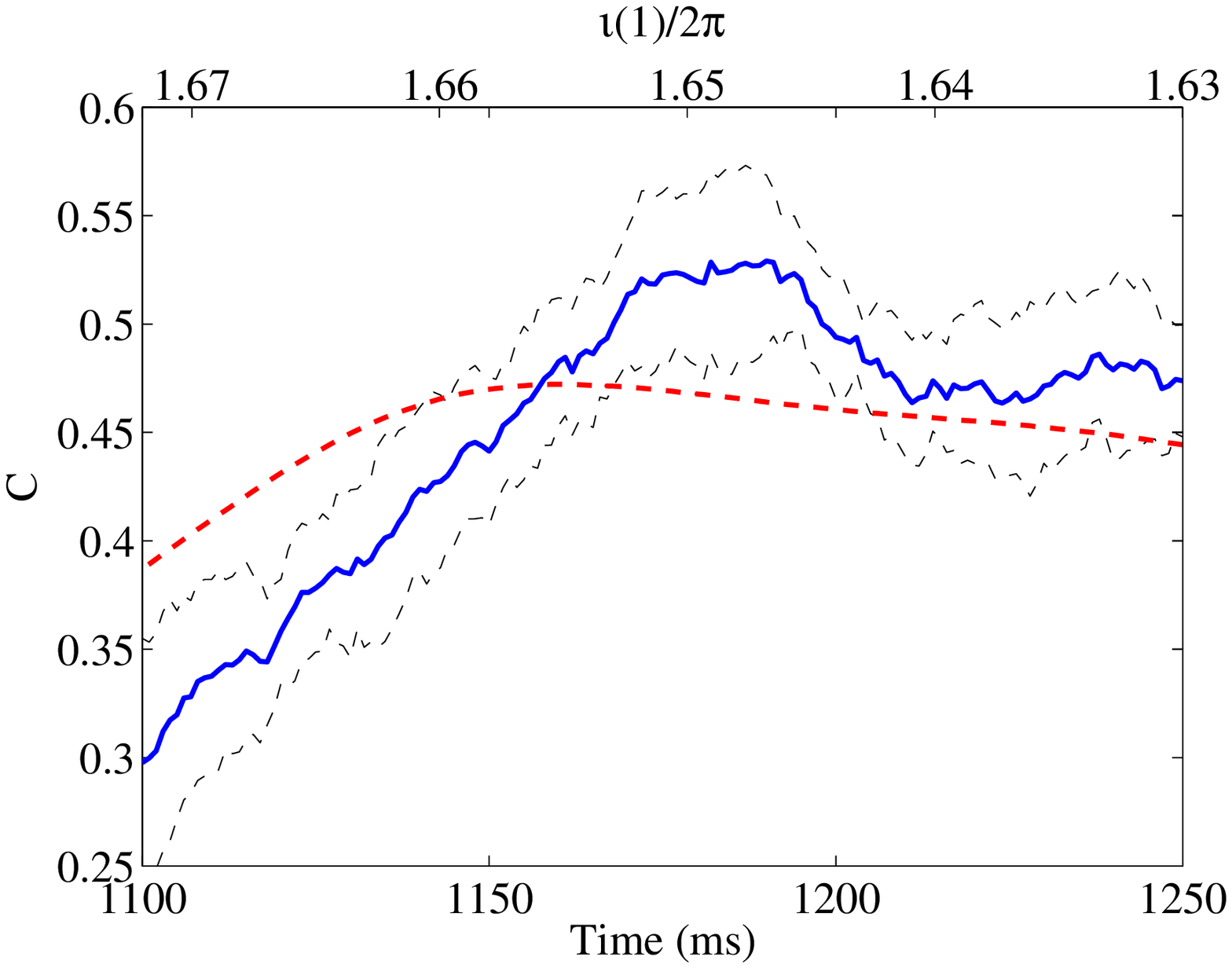}
  \includegraphics[trim=0 0 0 0,clip=,width=7.5cm]{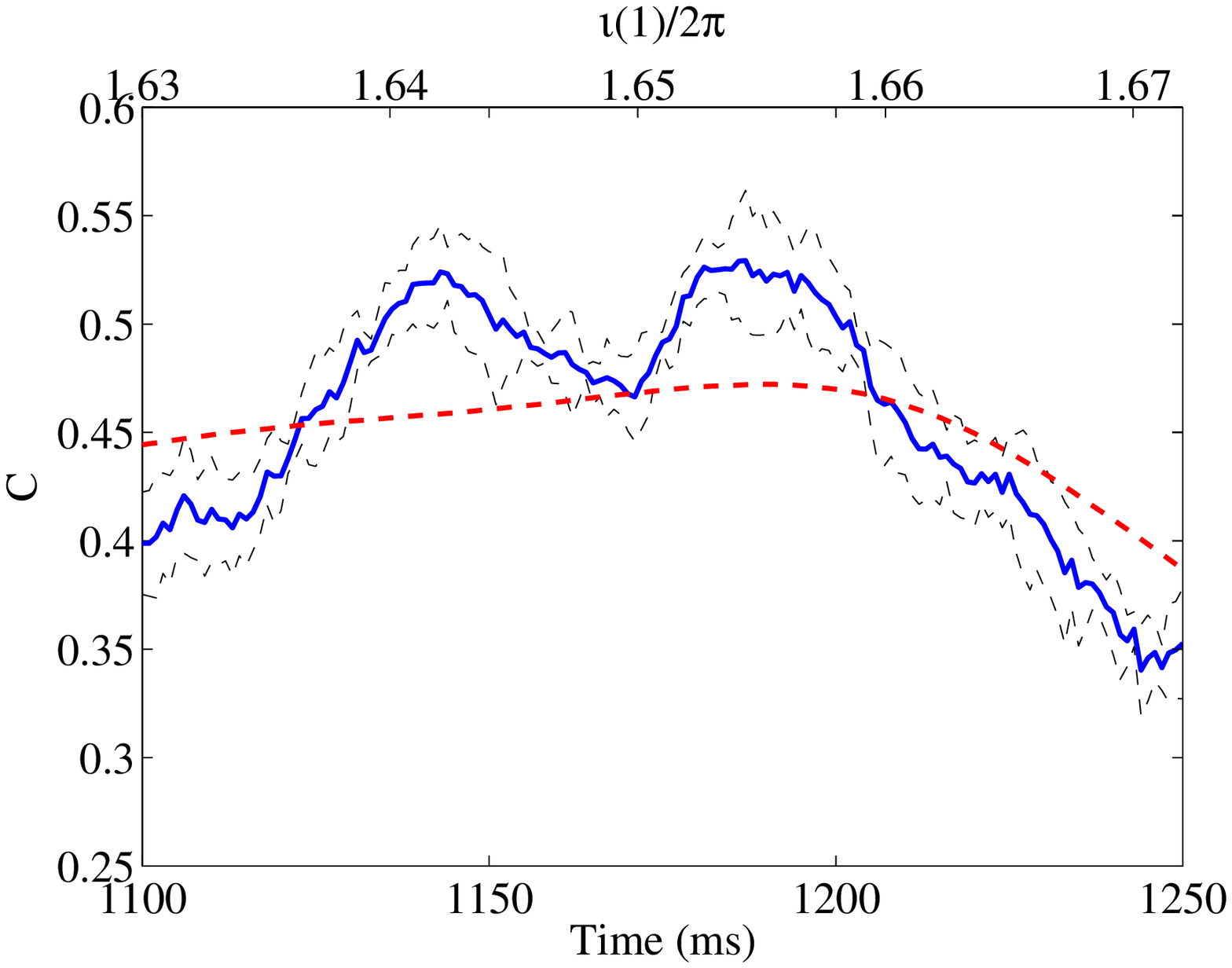}
\caption{\label{corr_scan1-2}Left: downward iota scan (scan 1). Right: upward iota scan over the same iota range (scan 2). Thick blue lines: experimental data. Thin black dashed lines indicate the range of shot to shot variation. Red dashed lines: model curves, Eq.~(\ref{C}), with $\lambda_\perp = 0.069 \pm 0.007$ m, $\lambda_{||} = 35 \pm 5$ m.}
\end{figure}

\begin{figure}\centering
  \includegraphics[trim=0 0 0 0,clip=,width=7.5cm]{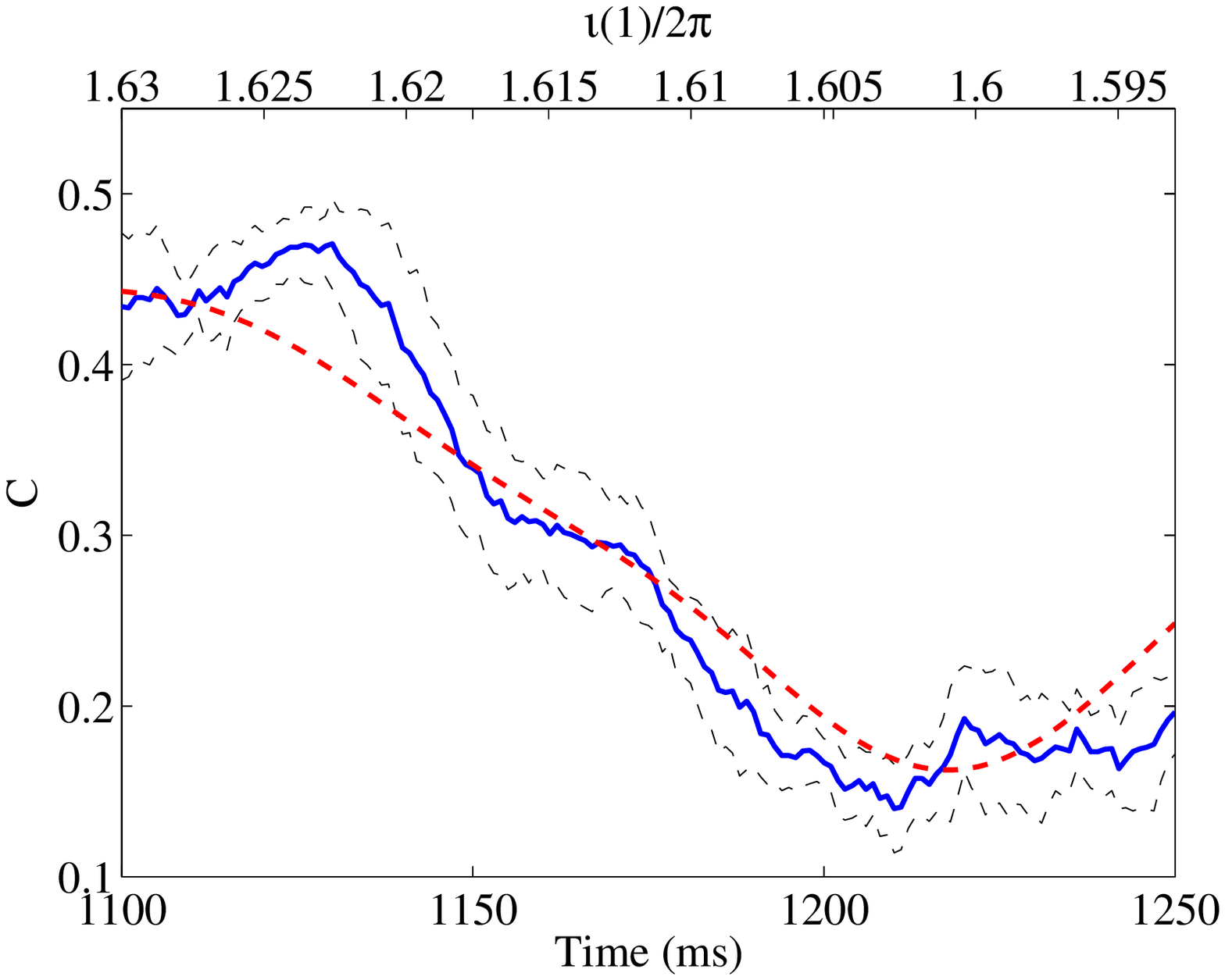}
  \includegraphics[trim=0 0 0 0,clip=,width=7.5cm]{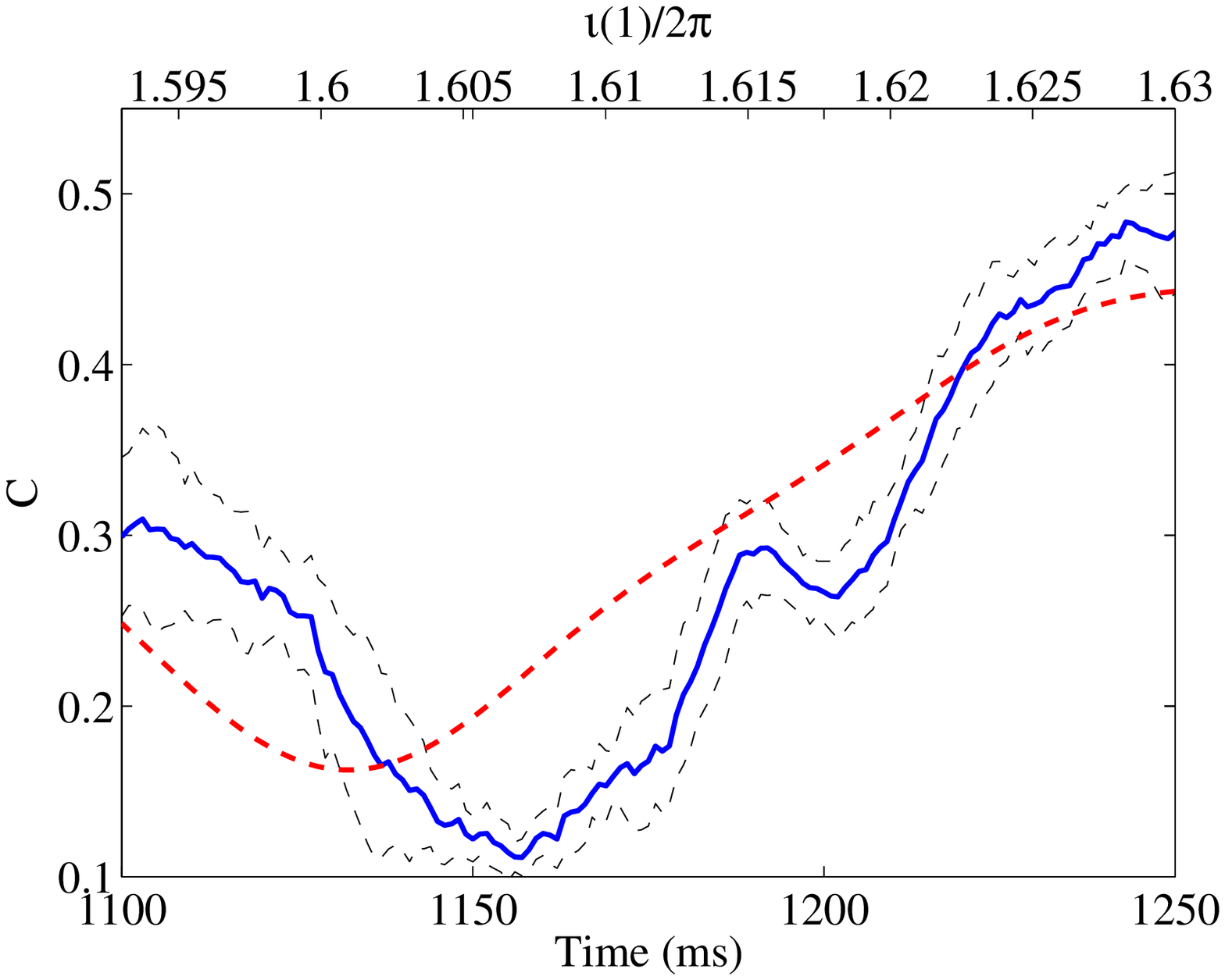}
\caption{\label{corr_scan3-4}Left: downward iota scan (scan 3). Right: upward iota scan over the same iota range (scan 4). Thick blue lines: experimental data. Thin black dashed lines indicate the range of shot to shot variation. Red dashed lines: model curves, Eq.~(\ref{C}), with $\lambda_\perp = 0.079 \pm 0.002$ m, $\lambda_{||} = 59 \pm 8$ m.}
\end{figure}

Fig.~\ref{corr_scan1-2} shows the mean experimental correlations for scans 1 and 2 and a joint fit with the function $C$, Eq.~(\ref{C}). 
Fig.~\ref{corr_scan3-4} shows the mean experimental correlations for scans 3 and 4 and a joint fit with the function $C$, Eq.~(\ref{C}). 
The shot to shot variation of the measured correlation is indicated in the figures, showing that the measured correlation is quite reproducible.

Although the model curves do not lie fully within the range of shot to shot variation indicated in the figures, it should be noted that this variation is not the same as the measurement error. The actual measurement error may be larger than the indicated variation, due to, e.g., systematic errors in the external coil currents, systematic errors in the probe position, contamination from minor mode activity, etc.
Furthermore, it may be that the model parameters $\lambda_\perp$ and $\lambda_{||}$ are not fixed, as assumed here for simplicity, but vary slightly as $\iota/2\pi$ is scanned.

\clearpage
\subsection{Long range correlation as a function of line average density}

\begin{figure}\centering
  \includegraphics[trim=0 0 0 0,clip=,width=7.5cm]{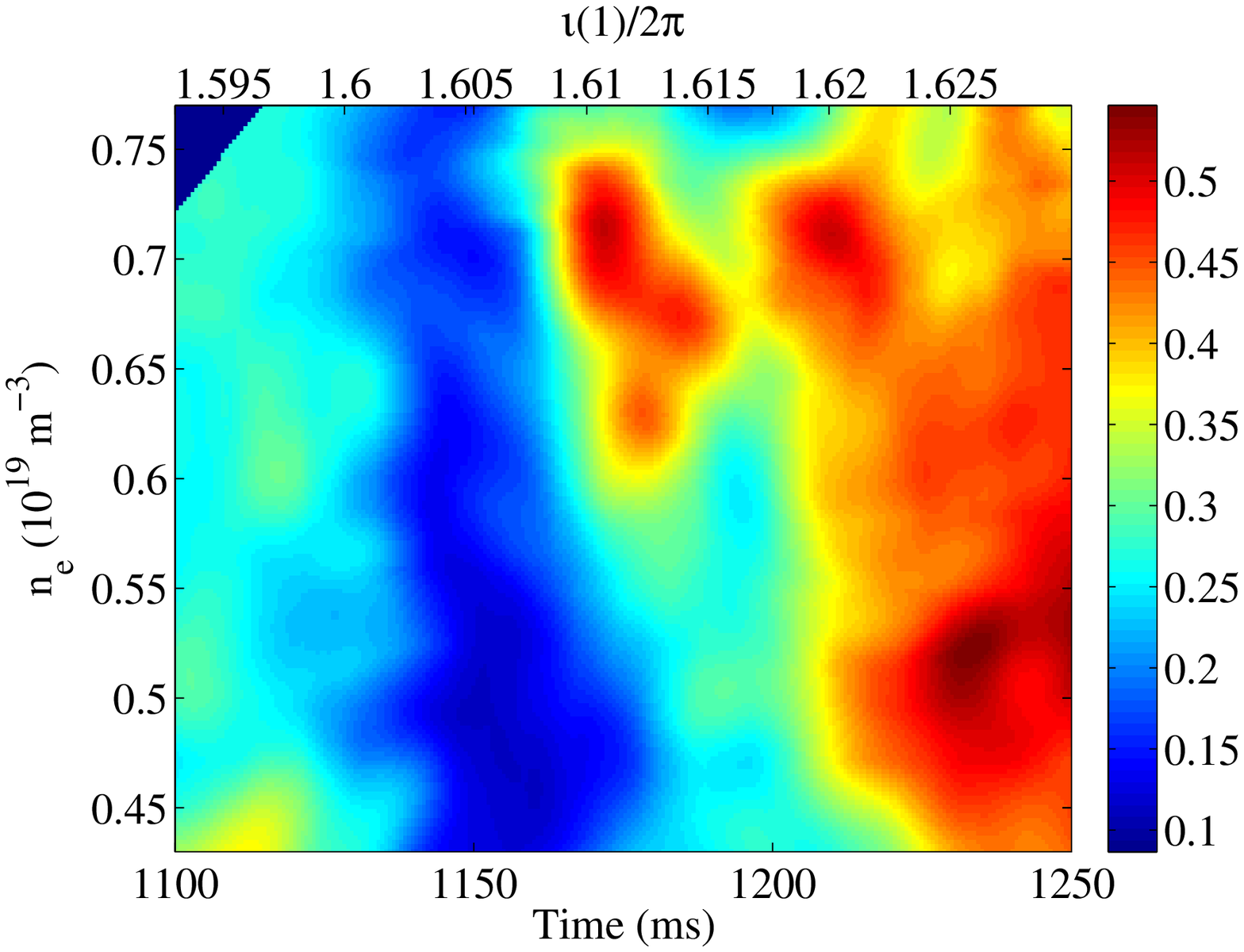}
  \includegraphics[trim=0 0 0 0,clip=,width=7.5cm]{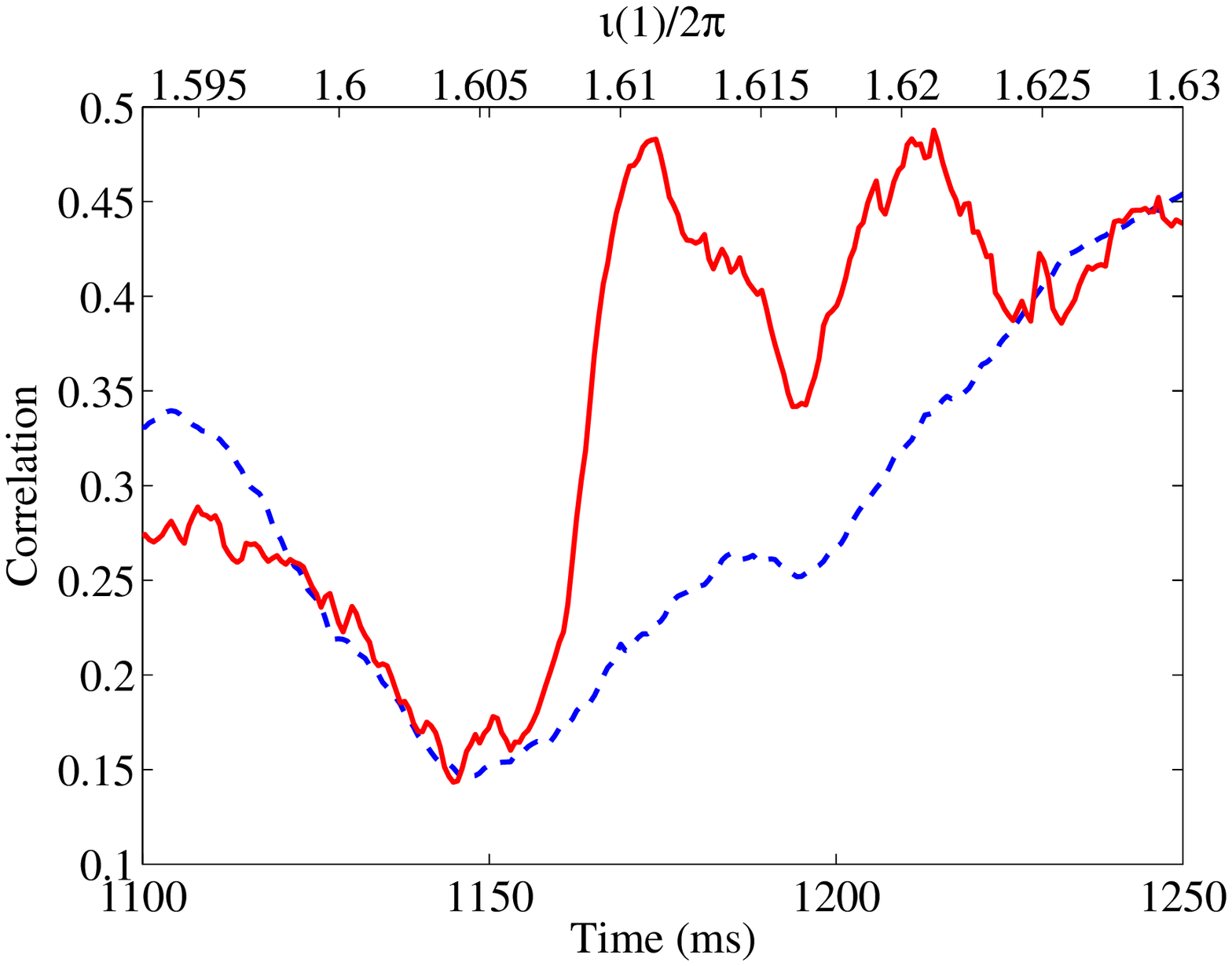}
\caption{\label{corr_scan5}Scan 5; upward iota scan (same as scan 4, cf.~Table \ref{table1}). 
Left: correlation versus time and line integrated density, $\overline{n_e}$. Color scale: probe correlation.
Right: mean correlation. Dashed blue line: mean correlation at low density, $\overline{n_e}< 0.6 \cdot 10^{19}$ m$^{-3}$.
Red continuous line: mean correlation at the critical density, $0.65 \le \overline{n_e} \le 0.73 \cdot 10^{19}$ m$^{-3}$.}
\end{figure}

After completion of Scan 4, this same scan was repeated in a set of 22 additional discharges in which the density was gradually raised on a shot to shot basis, while attempting to keep the density constant within each shot.
The resulting evolution of the correlation with time and density is displayed in Fig.~\ref{corr_scan5}.
At low densities, the temporal evolution of the correlation is nearly identical with the evolution shown in Fig.~\ref{corr_scan3-4} (right).
At higher density values, an additional correlation is observed, mainly in the time window $1170 < t < 1220$ ms and density values 
in the range $0.65 < \overline{n_e} < 0.73 \cdot 10^{19}$ m$^{-3}$. 

\subsection{Radial correlation}

Using five radially spaced Langmuir probe pins on the reference probe D, with a radial spacing of 5 mm and measuring floating potential, we also determined the radial correlation length $\lambda_r^D = 6.7 \pm 0.3$ mm. 
The radial correlation length obtained from six radially spaced pins with a radial spacing of 3 mm on probe B yielded $\lambda_r^B = 5.2 \pm 0.2$ mm.
One observes $\lambda_r^D > \lambda_r^B$, which can in part be explained by the larger flux expansion of the vacuum magnetic configuration at the position of probe D: $\lambda_r^B/\lambda_r^D = 0.78 \pm 0.5$ while $(\nabla \psi^B/\nabla \psi^D)^{-1} = 0.6$.
The radial correlation length varied very little during the four $\iota/2\pi$ scans of Table~\ref{table1}.

\begin{figure}\centering
  \includegraphics[trim=0 0 0 0,clip=,width=10cm]{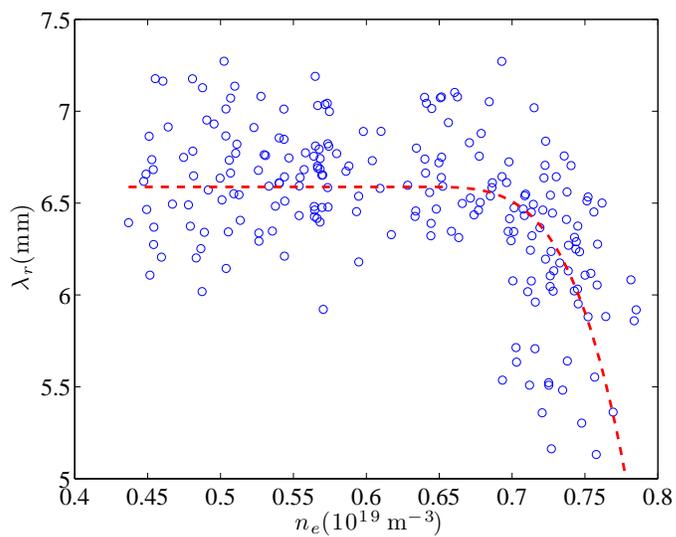}
\caption{\label{lambdar}Scan 5. Radial correlation length, $\lambda_r^D$, versus the local mean value of the line integrated density, $\overline{n_e}$. The line shown is meant to guide the eye.}
\end{figure}

The radial correlation was also monitored during the final scan at gradually increasing density.
Fig.~\ref{lambdar} shows that $\lambda_r^D$ remained very constant for $\overline{n_e} < 0.68 \cdot 10^{19}$ m$^{-3}$, and then dropped sharply (to about 5 mm) for higher density values.

\clearpage
\section{Discussion and conclusions}

In the specific experimental conditions of the four $\iota/2\pi$ scans (low density ECRH plasmas with tight control of the magnetic configuration), the highly reproducible correlation between two remote Langmuir probes seems to be well described by a simple model that assigns a correlation ellipsoid to a reference probe, with a long correlation length along the field line and a short correlation length in the perpendicular (mainly poloidal) direction.
The mentioned specific conditions were chosen to avoid driving strong global (MHD) modes, as well as zonal flows associated with the well-studied confinement transition at TJ-II (associated with a reversal of the edge radial electric field, $E_r$, and typically occurring at $\overline{n_e} \simeq 0.6 \cdot 10^{19}$ m$^{-3}$)~\cite{Hidalgo:2004,Guimarais:2007}.

The correlation lengths obtained by fitting a simple correlation model to the low density data, $\lambda_\perp \simeq 7$ cm and $\lambda_{||} \simeq 40-50$ m, are in accordance with expectations based on drift wave turbulence models~\cite{Chen:1984} and consistent with some earlier studies~\cite{Bengtson:1998,Thomsen:2001}.
The fact that the coherence is strong but not associated to a single frequency (Figs.~\ref{spec_scan1-2} and \ref{spec_scan3-4}) is consistent with the assumption that the correlation arises due to intermittent turbulent structures.
The perpendicular correlation length $\lambda_\perp$ is larger than the radial correlation length $\lambda_r \simeq 0.5-1$ cm, the latter being consistent with earlier studies~\cite{Pedrosa:2011,Silva:2011}.
The poloidal correlation length can be associated with a poloidal mode number $m = 2\pi \rho a / 4 \lambda_\perp \simeq 4-5$, possibly consistent with the presence of the 8/5 rational surface in the plasma edge region.

Fig.~\ref{corr_scan1-2} and \ref{corr_scan3-4} appear to show a small systematic time (or $\iota/2\pi$) shift between the modeled and measured correlation. 
In Fig.~\ref{corr_scan1-2}, the measured correlation appears to correspond to a slightly lower value of $\iota/2\pi$ than the calculated correlation, while in Fig.~\ref{corr_scan3-4}, the measured correlation appears to correspond to a slightly higher value of $\iota/2\pi$. 
Possibly, this shift can be explained by the presence of the $n/m = 8/5$ rational surface, as follows.
When the rotational transform at the probe position is approximately equal to $\iota(\rho_{B,D})/2\pi = 8/5$, the edge value of the rotational transform is $\iota(1)/2\pi \simeq 1.635$  (cf.~Fig.~\ref{iotascan}), i.e.,~very close to the lower limit of scans 1 and 2, and to the upper limit of scans 3 and 4.
The time (or $\iota/2\pi$) shift can therefore be explained if the measured correlation is not merely due to turbulence associated with the precise value of $\iota/2\pi$ corresponding to the probe position, but due to turbulence in a small radial range around the probe position, while the weight of the turbulence at the 8/5 surface contributing to the correlation is higher than weight of the turbulence at other values of $\iota/2\pi$ (or corresponding $\rho$ values).
This seems reasonable in view of the fact that any turbulent structures in the process of formation will have less tendency to be `smeared out' over the flux surface at a rational surface than at a non-rational surface.
This enhanced stability of turbulent structures on rational surfaces means that the turbulence corresponding to a rational surface will contribute with a relatively high weight to the measured correlation, thus producing the observed shift of the correlation.

In a final experiment, the probe correlation was monitored as the density was slowly raised.
Fig.~\ref{corr_scan5} summarizes the result. 
The background correlation due to turbulent structures seems to be essentially independent of the value of $\overline{n_e}$, but an {\it additional} correlation appears around the critical density $\overline{n_e} \simeq 0.65-0.73 \cdot 10^{19}$ m$^{-3}$, associated with a specific time window (i.e., value of $\iota/2\pi$).
This result is consistent with Fig.~4 of Ref.~\cite{Pedrosa:2008} (a density scan performed at $\iota(1)/2\pi = 1.653$).
In this previous work, this additional correlation was ascribed to the formation of zonal flows associated with a confinement transition.
After the transition has taken place (at the highest values of the density studied here), a shear flow layer and concomitant radial electric field, $E_r$, is established, suppressing turbulence and correlation and enhancing profile gradients.
This interpretation is consistent with the rather sharp drop of the radial correlation length observed here, presumably due to the impact of the sheared flow on the radial extension of the turbulent structures (cf.~Fig.~\ref{lambdar}).

The present work thus provides clear evidence that the observed effective correlation between remotely placed measurement systems (here: Langmuir probes) may be due to several mechanisms, operating simultaneously. 
The basic drift-wave model correlation mechanism appears to provide a correlation `baseline' on top of which the correlations due to other mechanisms, such as zonal flows, are mounted.
Consequently, in general one should be careful not to interpret a detected long-range correlation as being due to a single mechanism, unless further work is performed to clarify the precise cause of the correlation.

\section*{Acknowledgements}
The authors would like to express their gratitude for continued support by the TJ-II team.
Research sponsored in part by DGICYT of Spain under project Nr.~ENE2010-18409, and in part by the Ministerio de Econom\'ia y Competitividad of Spain under project Nr.~ENE2012-30832.


\section*{References}



\end{document}